 \documentclass[alpha-refs]{wiley-article}
\usepackage{graphicx}
\usepackage{subfig}
\usepackage{siunitx}
\usepackage{booktabs}
\title{High-Performance Statistical Computing (HPSC): Challenges, Opportunities, and Future Directions}
\author[1,\authfn{1}]{Sameh Abdulah}
\author[2,\authfn{1}]{Mary Lai O. Salva\~{n}a}
\author[3]{Ying Sun}
\author[1]{David E. Keyes}
\author[3]{Marc G. Genton}

\contrib[\authfn{1}]{These authors contributed equally to this work.}

\affil[1]{Applied Mathematics and Computational Science Program, King Abdullah University of Science and Technology, Thuwal 23955-6900, Saudi Arabia}
\affil[2]{Department of Statistics, University of Connecticut, Storrs, CT 06269-4120, USA}
\affil[3]{Statistics Program, King Abdullah University of Science and Technology, Thuwal 23955-6900, Saudi Arabia}

\corraddress{Marc G. Genton, Statistics Program, King Abdullah University of Science and Technology, Thuwal 23955-6900, Saudi Arabia}
\corremail{marc.genton@kaust.edu.sa}

\fundinginfo{King Abdullah University of Science and Technology}

\runningauthor{S. Abdulah et al.}

\begin{document}

\begin{frontmatter}
\maketitle

\begin{abstract}
We recognize the emergence of a statistical computing community focused on working with large computing platforms and producing software and applications that exemplify high-performance statistical computing (HPSC). The statistical computing (SC) community develops software that is widely used across disciplines. However, it remains largely absent from the high-performance computing (HPC) landscape, particularly on platforms such as those featured on the \\www.top500.org or Green500 lists. Many disciplines already participate in HPC, mostly centered around simulation science, although data-focused efforts under the artificial intelligence (AI) label are gaining popularity. Bridging this gap requires both community adaptation and technical innovation to align statistical methods with modern HPC technologies. We can accelerate progress in fast and scalable statistical applications by building strong connections between the SC and HPC communities. We present a brief history of SC, a vision for how its strengths can contribute to statistical science in the HPC environment (such as HPSC), the challenges that remain, and the opportunities currently available, culminating in a possible roadmap toward a thriving HPSC community.

% Please include a maximum of seven keywords
\end{abstract}
\end{frontmatter}

%%%%%%%%%%%%%%%%%%%%%%%%%%%%%%%%%%%%%%% Introduction
% 1-Overview of statistical computing (One paragraph)
% 2- Importance of high performance (HPC) in modern stats (two paragraphs)
% 3- Underutilization of MPI+X in Statistical Computing (One Paragraph)
% 4- Scope of the survey (One Paragraph)

\section{Introduction}   
% 1-Overview of statistical computing (One paragraph)
Statistical computing (SC) is a foundational discipline that aims to merge statistical theory with computational techniques to turn data into actionable insights~\citep{kennedy2021statistical}. At its core, it involves designing and implementing novel algorithms, simulations, and models to solve complex problems across various domains, including climate science, economics, and machine learning. Unlike general-purpose computing, statistical computing is grounded in data analysis, probabilistic reasoning, and statistical inference, drawing heavily on probability theory, statistics, and numerical methods~\citep{givens2012computational}. As data continues to scale in both size and complexity, statistical computing has become indispensable, not only for managing large datasets but also for enabling predictive modeling and informed decision-making in real-time. Programming languages such as R~\citep{ihaka1996r}, Julia~\citep{bezanson2017julia}, and Python~\citep{van1995python} have played a critical role in this evolution, empowering researchers to rapidly post-process, analyze, and visualize data. Together, these tools have made SC a central engine of modern data science.

% 2- Importance of high performance (HPC) in modern stats (two paragraphs)
At the same time, the field is facing a shift that is both technical and cultural: the convergence of SC with high-performance computing (HPC). The rapid growth of data and demand for more sophisticated analytics has pushed traditional, single-processor workflows to their limits. In response, SC is moving to distributed and parallel computing frameworks that leverage the massive scale of modern HPC architectures. To recognize this transition, we introduce the term \textit{high-performance statistical computing (HPSC)}, which formalizes the critical intersection where advanced statistical methods meet cutting-edge computational infrastructure. HPSC is not simply a technical move. It indicates a fundamental shift in the conceptualization and development of scalable solutions to statistical problems. Scientific applications are already extending the limits of HPC capabilities, particularly in areas involving large-scale simulations, uncertainty quantification, and real-time inference. Recognizing HPSC as a separate domain enables us to articulate its challenges more effectively, tap into its unexploited potential, and encourage deep collaboration between the HPC and SC communities.

% 3- Underutilization of MPI+X in Statistical Computing (One Paragraph)
The advancement of HPSC relies on a foundation of interdisciplinary collaboration, which brings together the strengths of modern computational science, statistics, and computer engineering. Progress in this emerging domain demands not only technical expertise in both statistics and computer science but also the ability to bridge these fields meaningfully. This integration requires a deep understanding of statistical literature, algorithmic design, parallel computing architectures, and hardware, as well as an understanding of domain-specific challenges. When these fields intersect, they enable the development of high-performance implementations that preserve statistical accuracy while fully leveraging the computational power of HPC platforms.  This collaborative approach will become increasingly critical to advancing the field as data volumes continue to grow and computational demands increase, driving the next wave of scalable, statistically sound solutions to meet the challenges of modern data analysis.

To date, a lot more work within the SC community has been invested in adopting dataflow technologies, i.e., frameworks such as Apache Spark~\citep{hariadi2020prediction}, Dask~\citep{rocklin2015dask}, or TensorFlow~\citep{abadi2016tensorflow}, that represent computations as directed acyclic graphs of operations on data than in exploring message passing interface (MPI)+X approaches, i.e., a hybrid parallel programming paradigm that combines MPI~\citep{gropp1999using} for distributed-memory communication with an additional model ``X” (such as OpenMP~\citep{chandra2001parallel} for multicore CPUs or CUDA for GPUs) for shared-memory or accelerator-level parallelism~\citep{MPI-X2021}, despite the latter being better suited for many of the most computationally demanding tasks in the field. There are multiple factors behind this trend. First, dataflow architectures are often perceived as more accessible, thanks to their relatively simple parallel execution models, in contrast to MPI-based systems, which require explicit message passing and distributed memory management. Such frameworks abstract away much of this complexity. This helps statisticians use them even if they do not fully understand parallel programming. Second, the need for interactive computing has strongly influenced choices. Tools like R and Python are deeply tied to exploratory, iterative analysis, where users develop models, visualize outputs, and debug interactively. Moreover, dataflow systems integrate more naturally into cloud-based and batch-processing environments, making them a compelling choice for both industry and large-scale analytics. Third, most statisticians and data scientists primarily work with multicore processors rather than large HPC clusters. This has led to a strong emphasis on shared-memory parallelism (e.g., OpenMP, threading in R and Python) rather than distributed memory paradigms, such as MPI. 

While dataflow technologies offer some distributed computing capabilities, they are often employed in a manner that still aligns with a multicore approach rather than fully leveraging the power of supercomputing. Even when statisticians engage with distributed computing, it is often through tools such as Dask~\citep{rocklin2015dask}, Ray~\citep{moritz2018ray}, or Spark, which offer a gentler learning curve and allow scaling across nodes without requiring low-level programming expertise. While these tools have clear advantages in usability and accessibility, we confirm that MPI+X remains significantly underutilized in SC. Supercomputing environments offer high efficiency, lower latency, and access to highly optimized numerical libraries. These advantages can substantially accelerate tasks such as large-scale simulations, Monte Carlo methods, and Bayesian inference. Fully embracing HPC technologies can unlock performance levels far beyond what current dataflow frameworks typically provide. In the following sections, we delve deeper into these challenges and present a roadmap for making MPI+X approaches more accessible, scalable, and impactful for the SC community.

%Scope of the survey (One Paragraph)
This paper attempts a comprehensive review of HPSC, examining the current landscape of its challenges, opportunities, and future directions in a field that is rapidly evolving. Our analysis includes recent advances in distributed computing frameworks, parallel processing algorithms, and modern hardware architectures, which have significantly expanded the capabilities of SC. This review focuses on how current statistical applications do not fully utilize the MPI+X approach, thereby missing out on significant performance gains. Through a systematic survey of the literature and prevailing practices, we identify key technological trends, methodological breakthroughs, and emerging research trajectories that are shaping the future of HPSC. The scope of this paper bridges both theory and practice, with a focus on how statistical methodology can effectively align with HPC infrastructure to meet today’s growing computational demands. Herein, we address critical issues such as scalability, reproducibility, and accessibility, highlighting strategies that have proven successful and pointing to areas where further innovation is needed. Our goal is to provide a valuable resource for researchers and practitioners working at the intersection of statistics and HPC, and to offer a forward-looking perspective on the evolving role of HPSC in data-intensive science.

This paper is organized as follows. Section 2 shows the historical development of HPC and SC, highlighting their progressive convergence. Section~\ref{sec:literature} surveys the literature, illustrating how HPSC is transforming various application domains. Section~\ref{sec:challenges} identifies key computational and methodological challenges in adapting statistical algorithms to modern HPC platforms. In Section~\ref{sec:opportunities}, we explore opportunities where HPC can enable scalable, precise, and energy-efficient statistical analysis. Section~\ref{sec:future} outlines future directions, focusing on specialized hardware, federated inference, standardization, and new algorithmic paradigms. Finally, Section~\ref{sec:roadmap} presents a roadmap for building a sustainable and inclusive HPSC community.

\section{HPC and SC: A History}
\label{sec:history}
In recent years, the evolution of computing hardware has entered a new phase, driven by the slowing of transistor scaling and the growing demand for high-performance processing. This marks a turning point from the earlier era when Moore’s Law reliably predicted rapid gains in transistor density. In 1965, Gordon Moore, co-founder of Intel, extrapolated four years of data to predict an exponential increase in the number of transistors per integrated circuit (Moore, 1965).  The ensuing decades bore him out with a doubling period of approximately 18 months. However, that pace has since slowed. By 2005, the rise in single-chip central processing unit (CPU) performance, powered by millions of transistors, began to face serious thermal and power limitations~\citep{sutter2005free}. In response, hardware designers shifted focus from increasing clock speeds to integrating multiple cores on a single chip, introducing the era of multicore processors. These architectures introduced multiprocessing execution, placing new demands on software to manage parallelism effectively. In parallel with the shift toward multicore CPUs, another revolution was quietly taking shape. At the end of the 20th century, NVIDIA introduced the first standalone graphics processing unit (GPU), the ``GeForce 256'', designed to accelerate gaming graphics~\citep{nvidia1999geforce}. What began as a specialized tool for rendering images quickly evolved into a powerhouse for general-purpose computing. With far greater parallelism and memory bandwidth than traditional CPUs, GPUs have become indispensable accelerators in modern HPC, at the expense of less memory per processor. Today, GPUs play a central role not only in gaming but also in high-performance applications, including deep learning, scientific simulation, and large-scale statistical inference. Modern HPC systems capitalize on this capability, combining GPUs with multicore CPUs to solve problems that exceed the limits of conventional computing. These platforms are specifically designed to address computational challenges that require vast processing power, high memory throughput, and low-latency communication. Supercomputers, the summit of HPC infrastructure, integrate thousands of CPUs and GPUs connected by ultra-fast interconnects such as InfiniBand \citep{grun2010introduction} and Slingshot \citep{de2020depth}. This tightly coupled architecture enables rapid coordination across processing units, allowing for the efficient execution of large-scale simulations and data-intensive workloads. Beyond traditional processors, modern HPC systems increasingly incorporate specialized accelerators, such as Tensor Processing Units \citep[TPUs,][]{jouppi2017datacenter} and Field-Programmable Gate Arrays \citep[FPGAs,][]{putnam2014reconfigurable}, to optimize specific tasks, including deep learning inference and cryptographic operations.

To maximize performance, many existing HPC environments rely on heterogeneous computing frameworks that integrate CPUs and GPUs into cooperative workflows. These systems excel at handling highly parallel workloads, often with dramatic gains in throughput and efficiency. Performance is commonly evaluated across several dimensions: FLOPS (Floating-Point Operations Per Second), which measures raw computational speed; memory bandwidth, which affects data transfer rates; a start-up time for an action that must be amortized over the amount of data transmitted or processed; and scalability and parallel efficiency, which reflect how well performance holds as the workload or system size increases; and energy efficiency.

%Since their inception, supercomputers have been closely serving military and defense applications~\citep{johnson1997supercomputing}. During World War II, high-speed computation emerged as a strategic asset. For instance, in the U.S., the development of ENIAC was driven by the need to automate artillery table calculations~\citep{metropolis1980man}, while Nazi Germany relied on the Enigma machine for encryption. In response, British cryptographers developed the Bombe and Colossus machines, which successfully broke German codes and contributed decisively to the Allied war effort. During the Cold War, supercomputing expanded into nuclear weapons design, aerospace development, intelligence processing, and code breaking.

\begin{figure*}
\centering
    \includegraphics[width=0.95\textwidth]{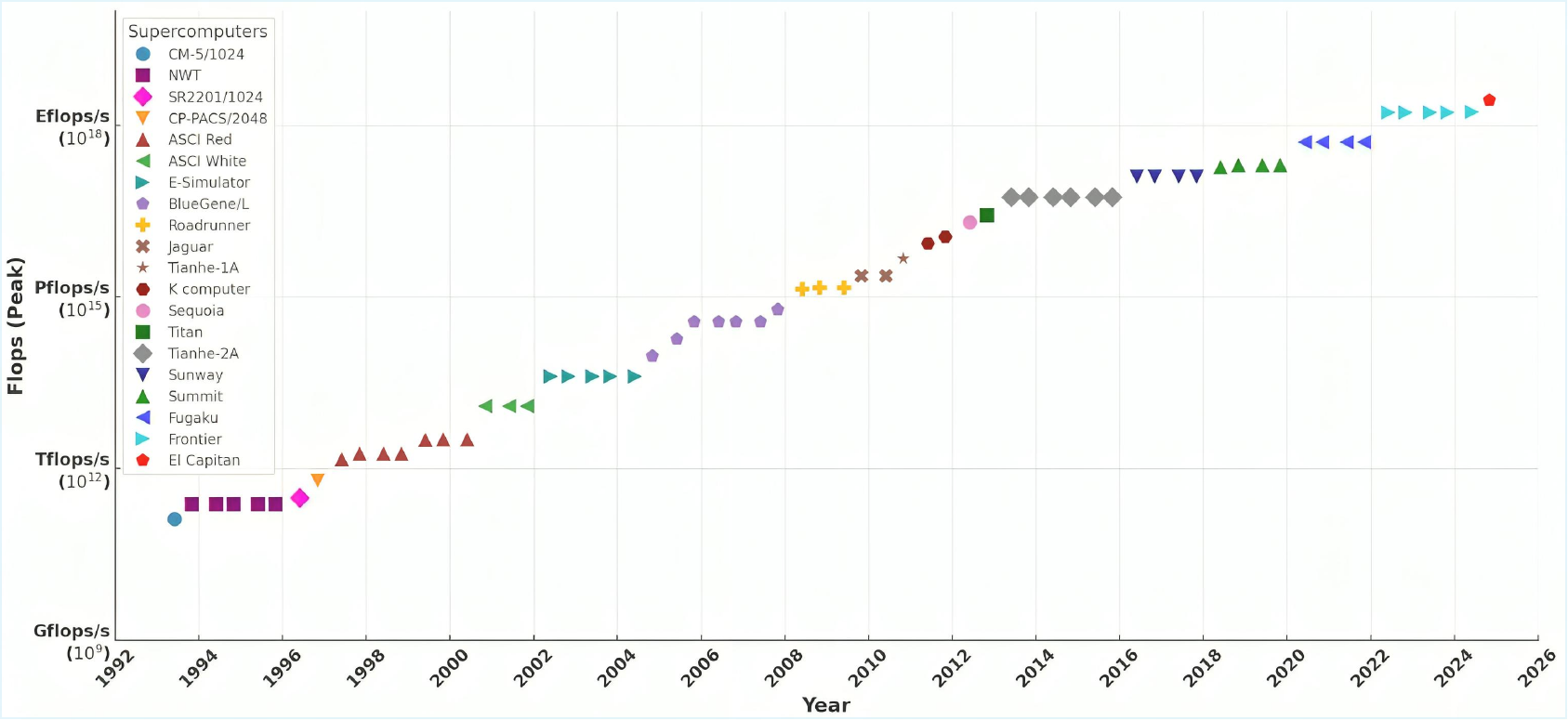} % Change filename
    \captionof{figure}{Evolution of Supercomputer Peak Performance Over Time: From CM-5 to El Capitan.}
    \label{fig:supercomputers}
\end{figure*}

The modern era of HPC began with the 1976 debut of the Cray‑1, which delivered roughly 133 MFLOPS. Progress accelerated quickly, producing more HPC machines. The Cray X-MP (1982) reached 800 MFLOPS, the Intel iPSC Hypercube (1985) surpassed 1.6 GFLOPS, the Cray Y-MP (1988) hit 2.67 GFLOPS, and the CM-5 (1991) achieved 131 GFLOPS~\citep{dongarra1990set}. These advances fueled breakthroughs across science, defense, and industry, introducing HPC as a foundational technology for large-scale computation. Over the past three decades, technologically leading nations have engaged in an ongoing race to develop and possess the world's fastest supercomputers, as represented in the biannual TOP500 rankings~\citep{Top500}. These machines have redefined what is computationally possible, powering breakthroughs in many applications, including physics, climate modeling, AI, and engineering. Their rise has been fueled by rapid innovations in CPU, GPU, and system architecture, often driven by advancements in the industry. Figure~\ref{fig:supercomputers} shows this trajectory from 1993-2024, highlighting the top-performing supercomputers at each point in time. Peak performance, expressed on a logarithmic scale in FLOPS, highlights the exponential growth trend that has characterized HPC.  From gigaflops in the early 1990s to exaflops today, the curve reveals a consistent doubling of computational power roughly every 1.5 to 2 years, mirroring Moore's Law and the scaling of parallelism. Some systems, such as ASCI Red, BlueGene/L, Summit, Fugaku, and most recently Frontier and El Capitan, mark significant inflection points in this evolution. If this trend can be sustained, zettaflop-scale systems are expected to emerge as early as the 2030s, introducing a new era of extreme-scale computing for many scientific fields.

%%%%%%%%%%%{TODO} Sameh: We need the PDF version

SC also has a long history dating back to the mid-20th century. One of the earliest documented statistical packages, the Bio-Medical Data Package (BMDP), was developed in the 1960s at the UCLA Health Computing Facility for IBM mainframes~\citep{dixon1983bmdp}. Designed for parametric and nonparametric statistical analyses, BMDP marked the first wave of statistical computing. The second wave emerged in the late 1960s and early 1970s with the release of SPSS (1968), Genstat (1970), and SAS (1971) \citep{nie1970SPSS}.  These packages introduced powerful tools such as regression, ANOVA, and cluster analysis to mainframe computing, revolutionizing statistical analysis in research, and were written in FORTRAN. {\color {black} The third statistical computing wave began} with the development of the S language in the 1970s, led by John Chambers and his team at Bell Labs \citep{chambers2008software}. Originally developed as a general-purpose statistical tool for academic users, S was later acquired by Insightful Corporation and rebranded as S-PLUS, which emerged as a leading commercial statistical software in the 1990s. Around the same time, Robert Gentleman and Ross Ihaka introduced R~\citep{ihaka1996r}, establishing a new standard for open, extensible, and efficient statistical computing. Their work, published in the Journal of Computational and Graphical Statistics, laid the foundation for modern statistical software and established key principles for efficient statistical computation that remain relevant today. The late 1990s and early 2000s saw the convergence of this foundation with advances in parallel architectures and algorithms, setting the stage for a new chapter. This paper introduces the fourth wave in SC, i.e., HPSC. Unlike previous waves, HPSC is not defined by a specific language or software, but by a broader shift toward scalable and parallel approaches that leverage the power of modern HPC systems.

Today, HPC includes a broad spectrum of technologies, most of which involve some form of parallel computing. Two dominant tracks have emerged: one rooted in the scientific community, the other in industry. In science, HPC is structured by national supercomputing centers focused on simulation science, typically leveraging the MPI+X programming model~\citep{MPIuse2020, MPI-X2021}. Interestingly, \cite{MPI-X2021} uses R for portions of their data analysis and graphics. In contrast, the business world has adopted large-scale data centers, driven by cloud computing, search engines, and e-commerce, which favor dataflow-based parallelism~\citep{MPIvsDataflow2018Fox}. These divergences reflect fundamental differences in application needs, as well as a degree of cultural and technical isolation between the two communities. At first glance, one might assume the dataflow model is better suited to statistical computing due to its focus on data processing. However, the reality is more complex and in many cases, the opposite holds true. A careful comparison of three widely used algorithms in computational statistics\footnote{Referred to as machine learning algorithms in the source, reflecting its computer science perspective.} shows that MPI-based implementations consistently outperform their dataflow counterparts, often by significant margins~\citep{MPIvsDataflow2018Fox}. The MPI model, grounded in mathematical structures such as matrix algebra, aligns naturally with the computational patterns of statistical analysis. While dataflow systems offer strengths in scalability and data management, especially in long-running services, they are optimized for use cases distinct from the demands of HPSC.

\section{HPSC in the Literature}
\label{sec:literature}
{\color{black} HPSC applies HPC to complex statistical problems, enabling massive data analysis, large-scale simulations, and models that were once infeasible. This convergence now benefits many fields. This section highlights areas where HPSC already delivers impact and continues to drive innovation.}

%%%%%%%%%%%%%%%%%%%%%%%%%%%%%%%%STOP
%%%%%%%%%%%%%%%%
\subsection{Climate Science}
HPSC has significantly advanced climate science by enabling scalable inference, prediction, and uncertainty quantification for massive spatio-temporal environmental datasets. In the literature, a notable contribution is the \textit{ExaGeoStat} framework~\citep{abdulah2018exageostat, abdulah2018parallel, abdulah2019geostatistical}, which introduced tile-based, architecture-agnostic solvers for dense and approximate Gaussian process (GP) likelihoods, supporting exact and approximate maximum likelihood estimation (MLE) for spatio-temporal data modeling on CPU-GPU clusters. {\color {black} Tile algorithms partition matrices into cache-sized tiles, enabling efficient multicore/GPU parallelism~\citep{gustavson2001new, buttari2009class}.} 

{\color {black} A key innovation in \textit{ExaGeoStat} is the integration of tile low-rank (TLR) approximations, where each matrix tile is individually compressed in low-rank~\citep{abdulah2018parallel}, with mixed-precision techniques, where each tile is computed at a precision level (FP64/FP32/FP16) chosen adaptively based on matrix norms or application-specific features~\citep{abdulah2019geostatistical,cao2022reshaping,cao2023reducing}. This approach reduces memory footprint and computational cost while preserving likelihood fidelity~\citep{cao2022reshaping}.} 

Later extensions to \textit{ExaGeoStat} pushed beyond Gaussian assumptions~\citep{mondal2022parallel, mondal2023tile}, added hybrid-precision solvers with dynamic scheduling via PaRSEC~\citep{abdulah2021accelerating}, and enabled containerized deployments across diverse HPC systems~\citep{abdulah2024portability}. These innovations enabled large-scale inference on existing supercomputers with notable speedups in modeling evapotranspiration fields. To improve usability, \textit{ExaGeoStat} was wrapped as the R package \textit{ExaGeoStatR}~\citep{abdulah2023large}, bridging high-performance modeling and the R computing environment. Within this ecosystem, \cite{salvana2021high, salvana2022parallel} developed massively parallel MLE solvers for multivariate and spatio-temporal GPs, leveraging distributed dense linear algebra and global parallel particle swarm optimization~\citep{schutte2004parallel} to support exascale kriging for climate and pollution data.  \cite{geng2023gpu, geng2025gpu} further enhanced performance by implementing a GPU-accelerated modified Bessel function, which is integrated into \textit{ExaGeoStat} for the efficient evaluation of the Mat\'ern covariance function. Other \textit{ExaGeoStat} features include convolutional neural networks-based partitioning of nonstationary spatial domains~\citep{nag2025efficient}, diagnostics for predictive efficiency in low-rank models~\citep{hong2021efficiency}, and large-scale simulation capabilities showcased in three international competitions~\citep{huang2021competition, abdulah2022second, hong2023third}, which benchmarked performance across a range of covariance structures and approximations. \textit{ExaGeoStat}'s contribution to HPSC in climate science was recognized when \cite{abdulah2024boosting} received the ACM Gordon Bell Prize for Climate Modeling.

Beyond \textit{ExaGeoStat}, HPSC-enabled tools have advanced climate modeling through GPU-accelerated profile likelihoods for dense spatial covariance matrices~\citep{xu2024profile}, decentralized low-rank inference~\citep{shi2025decentralized}, sparsity‑discovering kernels~\citep{noack2023exact}, parallel particle-based Bayesian estimation~\citep{katzfuss2017parallel}, and parallel Markov Chain Monte Carlo (MCMC) techniques~\citep{solonen2012efficient}. {\color {black} Moreover,~\cite{fattah2025stiles} introduced sTiles, a GPU-accelerated framework for sparse Cholesky factorization of structured matrices, with a particular focus on block arrowhead types commonly arising in Bayesian inference and scientific computing.}

A GPU-powered neural network has also demonstrated significant potential for fast estimation of spatial covariance parameters \citep{gerber2021fast}. These methods address scale and communication bottlenecks while supporting robust, interpretable inference~\citep{mann2020nascent}. Additional contributions include an out-of-core GPU-based geostatistical interpolation~\citep{allombert2014out}, a parallel kriging technique based on the k-d tree method~\citep{wei2015kd}, and GPU-accelerated likelihood estimation software in R. Efficiency improvements in accelerating Cholesky decomposition using a mixed-precision approach~\citep{cao2023reducing}, GPU implementation of Vecchia approximations~\citep{pan2024gpu, james2024implementation, pan2025block, pan2025scaled}, and parallel approaches to inference of spatial extremes~\citep{castruccio2016high} further illustrate how HPSC reshapes both methodology and practice in climate science. 

\subsection{Geoscience}
HPSC has also become a driving force in geoscience, unlocking scalable modeling, simulation, and inference across massive spatial datasets. Traditional geostatistical methods, such as kriging, which have long been limited by their $\mathcal{O}(N^3)$ computational cost, are now being reimagined through GPU-accelerated and parallel frameworks. Substantial speedups, often exceeding 18X, have been achieved using CUDA for kriging~\citep{hu2016using,liu2022accelerating, de2022towards}, while OpenCL-based architectures offer similar performance gains~\citep{huang2016opencl}. Hybrid CPU-GPU systems have further extended acceleration to adaptive local kriging for earthquake studies~\citep{chang2018gpu} and the construction of LiDAR-derived digital elevation models~\citep{danner2012hybrid}. Delaunay tetrahedra~\citep{yao2015fast}, smoothing techniques, such as Savitzky-Golay filtering~\citep{de2022gpu}, and massively parallel GP regression~\citep{gramacy2014massively, krasnosky2022massively}, performed in GPUs, continue to drive these gains. Beyond interpolation, simulation and surrogate modeling have emerged as key application areas. For instance, \cite{gopinathan2021probabilistic} developed a multi-threaded emulation platform for tsunami scenario generation, while \cite{rouholahnejad2012parallelization} proposed a parallel processing
software that supports the hydrologic simulator SWAT (Soil Water Assessment Tool). Generalized likelihood estimation methods for hydrological inference~\citep{yin2020parallel} and fast spatio-temporal weighted regression algorithms~\citep{que2021parallel} have both been deployed on MPI clusters, fostering an ecosystem of scalable statistical methods. Visualization and simulation tools have also evolved, leveraging GPU rendering~\citep{heitzler2017gpu}, edge/cloud architectures~\citep{mudunuru2024perspectives}, and task-based schedulers~\citep{nesi2022multi} to handle dynamic, high-resolution geoscientific workflows. Applications span weather prediction \citep{muller2019escape}, nuclear radiation modeling~\citep{xiao2025application}, atmospheric optics~\citep{russkova2018monte}, and air pollution modeling~\citep{molnar2010air}. Furthermore, seismic hazard modeling~\citep{li2025spatially}, GPU-based surface reconstruction~\citep{yan2015speeding, yan2016high}, and gap-filling using the modified planar rotator method~\citep{lach2024fast} highlight the breadth of geoscience applications now running at scale. Underlying these achievements is a growing recognition that statistical methods, hybrid hardware, and adaptive software must co-evolve. Foundational studies on the usage of heterogeneous computer architectures for geoscientific datasets~\citep{prasad2015vision, prasad2017parallel} reinforce this message. Taken together, these efforts establish HPSC not just as an enabler, but as an essential computational backbone for next-generation geoscientific modeling and discovery.

\subsection{Genomics and Bioinformatics}
HPSC has become indispensable in genomics and bioinformatics, enabling scalable, reproducible inference across massive and heterogeneous biological datasets. The integration of distributed architectures and optimized numerical solvers has transformed phylogenetics, genome-wide association studies (GWAS), and early disease detection. In phylogenetics, parallel implementations such as TREE-PUZZLE~\citep{schmidt2002tree}, fastDNAml~\citep{stewart2001parallel}, and RAxML~\citep{stamatakis2008rapid,stamatakis2008efficient} laid the groundwork for likelihood-based inference at scale. These were followed by GPU-accelerated frameworks, such as BEAGLE~\citep{suchard2009many, ayres2012beagle, ayres2019beagle}, which are now central to Bayesian phylogenetic engines like MrBayes~\citep{pratas2009fine, ayres2017heterogeneous} and RevBayes~\citep{smith2024bayesian}. Recent tools, such as CMAPLE~\citep{trong2024cmaple}, push further, enabling pandemic-scale inference through memory-efficient likelihood evaluation and massively parallel topology search. In GWAS, performance breakthroughs have been achieved through the use of memory-efficient implementations and distributed matrix operations. Tools such as PBOOST~\citep{yang2015pboost}, rMVP~\citep{yin2021rmvp}, SAIGE-GPU~\citep{rodriguez2024gwasgpu}, BLUPmrMLM~\citep{li2024blupmrmlm}, MPH~\citep{jiang2024mph}, and Quickdraws~\citep{loya2025scalable} offer substantial speedups over legacy pipelines. At the extreme scale, \cite{ltaief2024toward} demonstrates mixed-precision Kernel Ridge Regression for multivariate GWAS on exascale architectures. Single-cell and spatial omics workflows are also increasingly HPSC-enabled. GPU-accelerated tools, such as SPAC~\citep{liu2025spac}, enable the scaling of spatial transcriptomics analysis to millions of cells. fastglmpca~\citep{weine2024accelerated} improves GLM-PCA for single-cell RNA sequencing by accelerating Poisson likelihood optimization. ParProx~\citep{ko2021computationally} enables penalized regression in ultrahigh-dimensional omics datasets. Other advances include parallel co-expression inference~\citep{shealy2019gpu}, GPU-based spike-triggered modeling~\citep{mano2017graphics}, a parallel minimum spanning tree construction algorithm~\citep{olman2008parallel}, and an HPC-AI hybrid framework~\citep {patil2024genome}, which achieves over a 200× speedup. Efficient scheduling across multi-GPU nodes~\citep{thavappiragasam2021addressing} and adaptive runtime optimization strategies~\citep{wang2024dnb} highlight the growing importance of workload balance and architectural efficiency. These innovations collectively demonstrate how HPSC is reshaping the field of genomics and bioinformatics applications.  As biological data continues to grow in scale and complexity, the integration of statistical modeling with hardware-optimized computation is enabling unprecedented levels of speed, performance, resolution, and insight.

\subsection{Physics and Astronomy}

HPSC has become a foundational enabler in physics, cosmology, and astrophysics, powering both simulation and inference for increasingly complex systems and datasets. GPU-accelerated pipelines lie at the heart of this transformation, driving advancements in a wide array of applications—from Monte Carlo simulations of spin models~\citep{barash2017gpu} to gravitational wave detection~\citep{talbot2019parallelized, wysocki2019accelerating, smith2020massively, bandopadhyay2024gpu, saltas2025emri_mc, jan2025adapting} and probabilistic programming for particle physics~\citep{baydin2019etalumis}. These capabilities are underpinned by GPU-native inference frameworks~\citep{fluke2011astrophysical, gindl2021gpu, dunn2022graphics}. In cosmology, HPC-compatible libraries such as CosmoPower~\citep{mancini2022cosmopower}, CONNECT~\citep{nygaard2023connect}, JAX-COSMO~\citep{campagne2023jax}, OL\'E~\citep{gunther2025ole}, and GLaD~\citep{wang2025gpu} enable rapid, high-dimensional posterior estimation with orders-of-magnitude improvements in runtime. Many of these tools incorporate likelihood inference~\citep{makinen2021lossless, ho2024ltu, piras2024future}, designed to leverage GPU acceleration for scalability. Large-scale simulations have also benefited dramatically from HPSC advances. Frameworks such as sCOLA~\citep{leclercq2020perfectly} and Farpoint~\citep{frontiere2022farpoint} leverage exascale computing to model cosmic structure formation, while GPU-accelerated linear algebra routines~\citep{rodrigues2014accelerating} and eigensolver libraries like ELPA~\citep{kus2019optimizations} serve core needs in quantum and accelerator physics. Emerging infrastructures are increasingly hybrid and cloud-enabled. HPC platforms that integrate CPU/GPU architectures~\citep{amadio2016electromagnetic, boyle2017accelerating} along with modular frameworks, such as $\Pi$4U~\citep{hadjidoukas2015pi4u}, CUDAHM~\citep{gindl2021gpu}, and AI4HPC~\citep{sarma2024parallel}, and HPC-portable languages, such as Julia in high-energy physics~\citep{stewart2025julia}, exemplify the fusion of high-performance engines with machine learning-driven inference. Despite these gains, challenges persist. Ensuring reproducibility across heterogeneous systems~\citep{jarp2012evaluation, tiwari2015understanding, krupa2021gpu}, standardizing inference error propagation, and generalizing emulators across physical domains remain open problems. Yet as HPSC continues to mature, it is becoming the statistical and computational backbone of physics-informed, simulation-driven discovery, transforming how uncertainty, complexity, and scale are addressed in modern physics.

\subsection{Economics}

HPSC is also rapidly reshaping economics, enabling researchers to tackle high-dimensional, nonlinear, and spatially structured models that were once computationally out of reach. In the literature, \cite{aldrich2011tapping} demonstrated how GPUs can be leveraged to solve dynamic equilibrium models, while \cite{creel2008multi} provided a broader tutorial on utilizing multicore architectures and parallel computing tools. Overviews by \cite{aldrich2014gpu} and \cite{villaverde_valencia18} helped codify GPU and parallel programming strategies tailored for economics. These foundations have since expanded across diverse econometric domains. Parallel Bayesian inference and adaptive sampling for econometric models have pushed the limits of inference in dynamic systems, as demonstrated by \cite{basturk2016parallelization}. Scalable toolkits, such as those introduced by \cite{chib2023dsge}, and the sequential Monte Carlo frameworks advocated by \cite{herbst2014sequential}, have improved tractability in complex posterior spaces. GPU-accelerated likelihood evaluations~\citep{white_porter14}, high-performance indirect likelihood inference methods~\citep{creel_zubair12}, and stochastic volatility models~\citep{gonzalez2025exact} underscore the power of hardware-aware design. Other key developments include embarrassingly parallel bootstraps~\citep{delgado2013embarrassingly}, and parallelizable linear transformation methods for mixed-frequency data~\citep{qian2016computationally}. Comparative studies of MPEC vs. NFXP~\citep{dong2022implementing} have clarified the computational trade-offs in structural estimation. In high-frequency settings, GPU-enabled Bayesian Hawkes process models~\citep{holbrook2021scalable, holbrook2022bayesian} have supported inference on spatio-temporal contagion data, while parallel Gibbs samplers have been applied to generalized autoregressive conditional heteroscedastic-intertemporal capital asset pricing models with extreme-value distributions~\citep{khanthaporn2023modelling}. Nonparametric and hierarchical methods continue to stretch computational resources. The \textit{np} package~\citep{hayfield2008nonparametric} enables flexible estimation via kernel methods but remains computationally intensive. Recent advances such as hierarchical geographically weighted regression with parallel backfitting~\citep{hu2024backfitting} and hierarchical panel data models for stochastic metafrontiers~\citep{amsler2023hierarchical} demonstrate how HPSC can address spatial and group heterogeneity at scale. Big-data-focused toolkits built on distributed computing frameworks, such as Apache Spark~\citep{hariadi2020prediction}, have also emerged as robust solutions for econometric modeling on massive datasets~\citep{bluhm2020spark}. Discrete games now leverage GPU-powered simulation and equilibrium search methods~\citep{chung2023efficient}. Parallel algorithms for solving finite mixture models~\citep{ferrall2005solving} and parameterized expectations in macroeconomics~\citep{creel2008using} further highlight the convergence of algorithmic innovation and hardware acceleration. Altogether, these developments showcase how algorithmic design, hardware optimization, and software engineering converge to redefine econometric methods in the era of complex data and real-time policy modeling.

\subsection{Finance}

\begin{table*}[h]
\caption{Recurring principles across HPSC applications.}
\centering
\footnotesize

\begin{tabular}{p{3.8cm} p{3.7cm} p{5.8cm}}
\toprule
\textbf{Principle} & \textbf{Domains} & \textbf{Representative Techniques/Tools} \\
\midrule
Approximations & Climate, Geoscience, Genomics, Finance & Tile/Block approximation, Vecchia, sketching, variational inference, approximate risk-neutral densities \\
Mixed-precision computing & Climate, Genomics, Finance & FP64/FP32/FP16 solvers, stochastic rounding, kernel ridge regression, energy-efficient likelihood evaluation \\
GPU/accelerator utilization & Physics, Climate, Finance, Genomics & CUDA/OpenCL kernels, cuBLAS/cuSPARSE, BEAGLE phylogenetics, GPU-accelerated MCMC, portfolio optimization, volatility modeling \\
%Parallel design patterns & All (including Finance) & Divide-and-Conquer, Task Parallelism, MapReduce, Streaming; Monte Carlo simulations, embarrassingly parallel bootstraps \\
Federated/distributed frameworks & Healthcare, Finance, Climate & Spark, Dask, Ray, Federated Gaussian Processes, Hadoop-based financial risk analysis \\
Software/hardware co-design & Climate, Genomics, Finance, Statistics & ExaGeoStat, RScaLAPACK, bigGP, pbdR, RCOMPSs; domain-specific HPC toolkits for finance (HXPY, stochastic simulation frameworks) \\
\bottomrule
\end{tabular}
\label{tab:domains}
\end{table*}

HPSC has revolutionized modern finance, enabling real-time analytics, high-dimensional inference, and complex simulation-based estimation across a wide range of applications, including portfolio optimization, volatility modeling, credit risk scoring, and derivative pricing. Early studies by \cite{zenios1999high} highlighted the role of HPC in accelerating Monte Carlo–based security pricing and Value-at-Risk models, while \cite{liyanage2017utilizing} demonstrated how such simulations could be radically optimized through vectorization and parallelization. Building on this, \cite{guo2023hxpy} introduced \textit{HXPY}, a high-performance data pipeline that leverages SIMD (Single Instruction Multiple Data) and CUDA to outperform legacy tools in financial time-series processing. Parallelization has also enhanced nonparametric risk-neutral density estimation, as seen in bandwidth-optimized approaches for options pricing~\citep{monteiro2023parallel}. Stochastic simulation frameworks for pricing and sensitivity analysis~\citep{dixon2012monte, dixon2014accelerating, kucherenko2023quasi} continue to benefit from high-throughput GPU implementations. Moreover, machine learning (ML) integration is a growing trend. CUDA-enabled ML pipelines have been applied to real-time stock forecasting~\citep{kumari2020cuda}, while parallelized ML approaches have accelerated credit risk scoring~\citep{hentosh2022ml}. Scalable Bayesian workflows for financial time series~\citep{casarin2016embarrassingly} push the limits of high-dimensional modeling.
Algorithmic innovations, such as parallel minorization–maximization algorithms for heteroscedastic regression~\citep{nguyen2016block}, have helped bridge the gap between statistical accuracy and computational speed. Recent works highlight the importance of the design trade-offs~\citep{inggs2013heterogeneous, jain2015accelerating, mudnic2017european}, emphasizing the need to balance model complexity, runtime constraints, and inference accuracy. Advances in massively parallel market environments for financial reinforcement learning~\citep{wang2025parallel} and high-fidelity calibration objectives for agent-based financial market simulations~\citep{wang2025alleviating} underscore the convergence of algorithmic modeling and high-performance engineering, while parallel Bayesian inference for stock price prediction on Apache Spark~\citep{hariadi2020prediction} further exemplifies this trend. As financial systems demand greater speed and precision, HPSC has become a critical pillar in the computational infrastructure of quantitative finance.

{\color {black}
\subsection{Recurring Principles Across HPSC Applications}
Across the different domains shown above, common principles highlight how HPSC is evolving. Approximation strategies, such as low-rank representations and randomized methods, can reduce computational and memory costs while maintaining statistical accuracy. Mixed-precision arithmetic and the use of GPUs and other accelerators are emerging as effective ways to balance performance with numerical stability across various fields. Parallel design patterns, including divide-and-conquer, task and data parallelism, and MapReduce, provide reusable frameworks for scaling statistical algorithms. Numerous studies have demonstrated the benefits of software and hardware co-design, where statistical methodologies are closely integrated with HPC libraries, containerization, and distributed frameworks. In conclusion, these recurring ideas highlight that while applications differ, HPSC is driven by a shared set of strategies that enable scalability, efficiency, and reproducibility across disciplines. Table~\ref{tab:domains} summarizes the different principles discussed in the above subsections.}

\section{HPSC Challenges}
\label{sec:challenges}
Leveraging HPC resources to accelerate statistical analysis presents challenges different from those of traditional computational science. Many statistical algorithms were initially designed for sequential execution and modest data sizes, making their transition to parallel, distributed, and heterogeneous architectures far from a straightforward process. These algorithms often exhibit complex data dependencies, irregular memory access, and heightened sensitivity to numerical precision, factors that complicate scaling on modern HPC platforms. Meeting these demands requires more than raw compute power. It requires algorithm redesign, minimizing data movement between computing units, ensuring numerical stability at scale, and building robust, portable, and reproducible software. In the following section, we outline the core challenges facing HPSC, including parallelization strategies, memory and data management, numerical precision, and implementation bottlenecks, with a focus on what is needed to achieve scalability and accuracy.

%%%%%%%%%%STOP
\subsection{Adapting Statistical Algorithms for Parallel Environments}

One of the main challenges in using HPC for statistical applications is that many fundamental algorithms were designed to be sequential by default. As a result, they often fail to scale effectively. Unlike simulation-based scientific codes, statistical routines frequently involve iterative, data-dependent logic that resists naive parallelization. The limited research on this issue focuses on the parallelization of statistical algorithms. Unlocking HPC performance thus requires more than just access to advanced hardware; it demands a redesign of the whole algorithm. This may involve restructuring existing methods to expose parallelism or developing new algorithms specifically designed for distributed environments and accelerators. This can be achieved by identifying independent computational tasks and blocks, optimizing data locality {\color{black} (i.e., keeping data physically close to where they are processed)} and communication patterns {\color {black}  (i.e., how parallel tasks, such as processes or threads, exchange data during computation)}, and balancing workloads across different processing units. Without such adaptations, statistical applications cannot fully exploit the power of HPC platforms.

The design of parallel algorithms always relies on the architecture of the target hardware. It also adds a layer of complexity for developers to consider. Thus, optimizing performance across various systems, such as multicore CPUs and GPU-accelerated clusters, requires careful alignment between the algorithmic structure and hardware characteristics. One of the central challenges is picking an appropriate parallel programming model and design pattern that fully exploits the available parallelism in the algorithm while minimizing overhead. {\color {black} Parallel programming models provide abstractions that define how computational tasks are structured, executed, and coordinated concurrently across multiple processing units. Examples include OpenMP~\citep{chandra2001parallel}, which enables shared-memory parallelism using threads; MPI~\citep{MPI-X2021}, which supports distributed-memory parallelism through message passing; and CUDA~\citep{nvidia2025cuda}, which allows fine-grained parallel programming on NVIDIA GPUs.} 

\begin{figure*}[htbp]
    \centering

    \subfloat[Data Parallelism\label{fig:data}]{
        \includegraphics[width=0.22\textwidth]{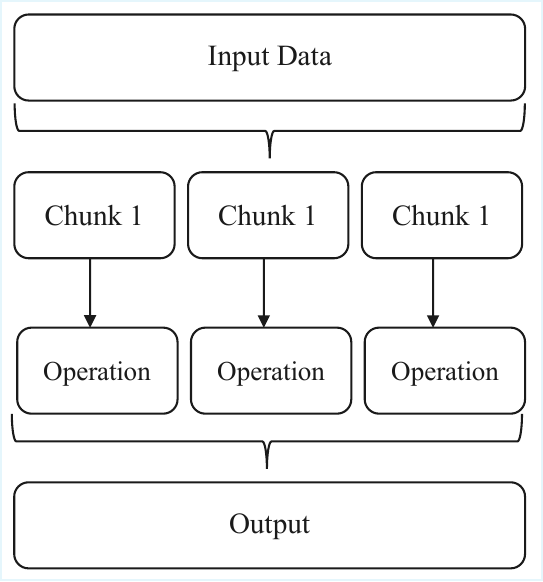}
    }
        \hfill
    \subfloat[Fork/Join\label{fig:forkjoin}]{
        \includegraphics[width=0.48\textwidth]{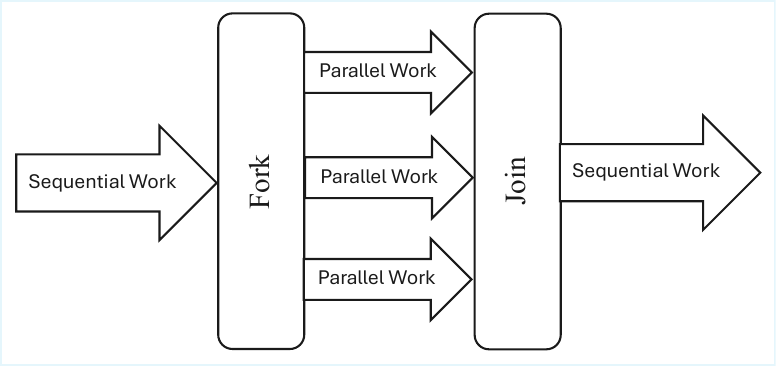}
    }
    \hfill
    \subfloat[Map/Reduce\label{fig:mapreduce}]{
        \includegraphics[width=0.22\textwidth]{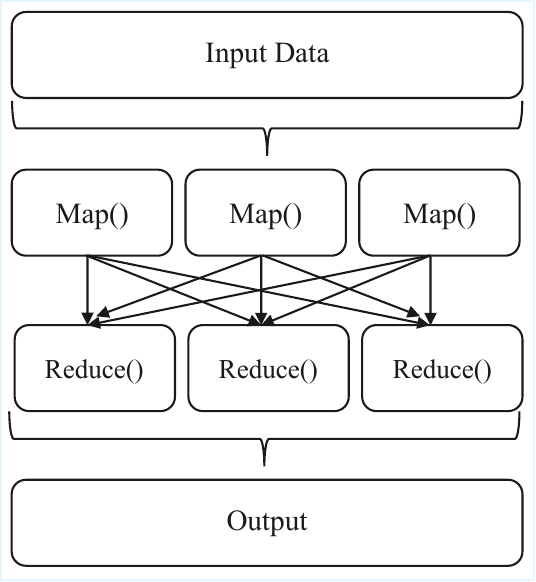}
    }

    \par\bigskip % Space between rows

    \subfloat[Divide and Conquer\label{fig:divideconquer}]{
        \includegraphics[width=0.4\textwidth]{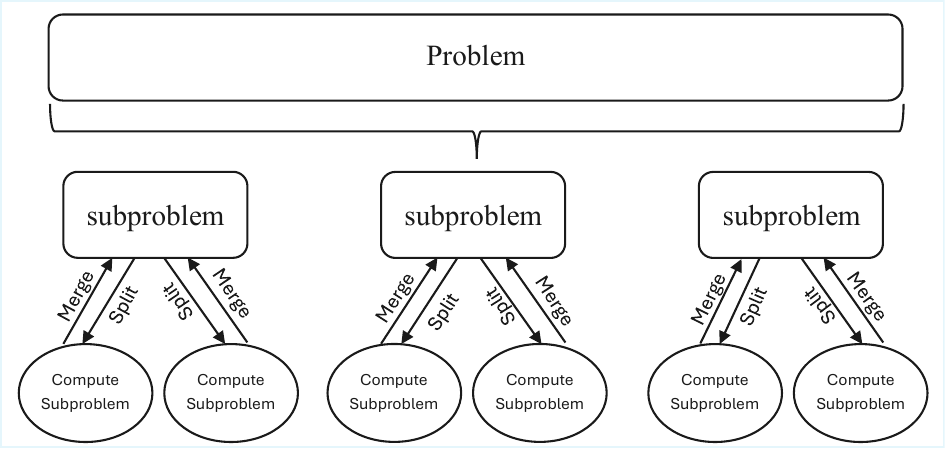}
    }
    \hfill
    \subfloat[Task Parallelism\label{fig:task}]{
        \includegraphics[width=0.24\textwidth]{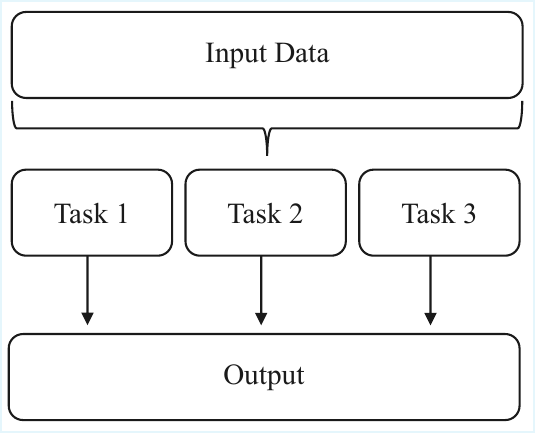}
    }
    \hfill
    \subfloat[Pipeline\label{fig:pipeline}]{
        \includegraphics[width=0.22\textwidth]{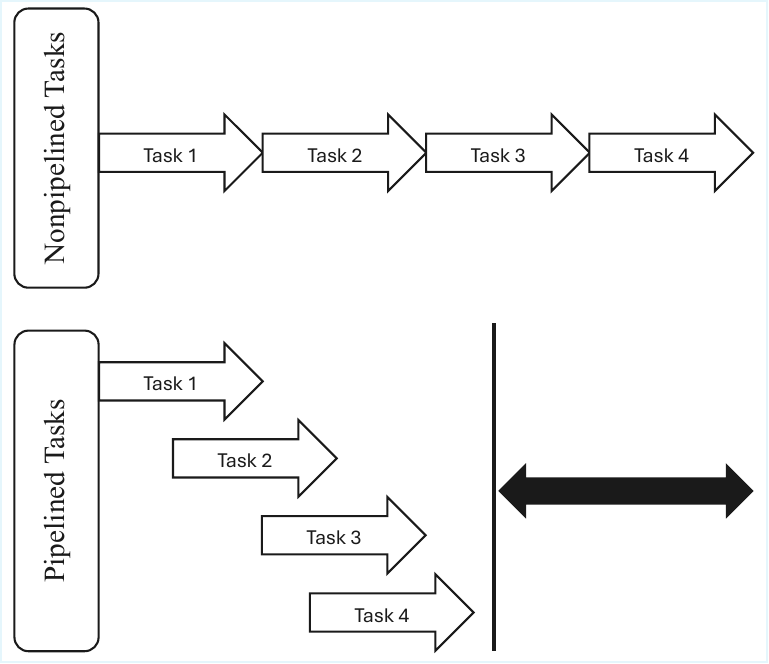}
    }

    \caption{Common Parallelization Patterns in HPC.}
    \label{fig:parallelpatterns}
\end{figure*}

Equally important is the choice of a suitable communication model, which governs how data and tasks are shared across processing units. Historically, the fork-join model has been the dominant paradigm, mainly in shared-memory settings. In this approach, execution begins sequentially until a parallelizable block is encountered, at which point multiple threads are spawned, typically mapped to individual cores, to perform concurrent work before merging back into the main thread. While simple and widely supported, this model can harm performance in certain scenarios. Specifically, it limits optimization opportunities related to data synchronization, thread scheduling, and cache efficiency, primarily when threads compete for shared memory resources. To overcome these bottlenecks, other algorithmic patterns are increasingly important, especially in the context of statistical computing, where data and computation structures are often irregular. Strategies such as Divide and Conquer, Task Parallelism, Data Parallelism, Map-Reduce, and Asynchronous Parallelism offer more flexible and scalable avenues for exploiting concurrency. Selecting the best pattern depends on the structure of the given task, the nature of data dependencies, and the characteristics of the underlying hardware. Incorporating these approaches early in algorithm design is essential for achieving high performance in HPC environments. Figure~\ref{fig:parallelpatterns} illustrates the six primary parallel computing patterns: Data Parallelism, Fork/Join, Map/Reduce, Divide and Conquer, Task Parallelism, and Pipeline Parallelism. Each subfigure illustrates a distinct execution structure, showcasing how computations or data flow are organized to exploit concurrency in modern high-performance and distributed systems. 

{\color{black} Different computing patterns suit different scenarios. Data parallelism is ideal when the same computation is run on many independent pieces, such as bootstrap resamples, cross-validation folds, or Monte Carlo draws. Map/Reduce is appropriate when ``group-by and aggregate" at scale is needed, e.g., summaries. A pipeline suits end-to-end workflows where data flow through stages. Fork/join fits phased methods that require a global combine step, e.g., computing partial gradients or sufficient statistics in parallel, then synchronizing to update a model. Divide and conquer is helpful when the problem naturally splits recursively, such as when partitioning the data or domain, solving subproblems, and merging the results. Task parallelism is more suitable for heterogeneous, dependency-driven work, such as hyperparameter searches, multi-model comparison pipelines, or sparse/graph computations.}

%%%%%%%%%%%%STOP
A growing number of studies in the literature illustrate how diverse parallel design patterns and programming models can be effectively applied to SC. For instance, \cite{hong2022divide} introduced the SOLID algorithm, a divide-and-conquer strategy for sparse predictive modeling and penalized regression in biostatistics. By coupling screening and one-step linearization, SOLID can achieve significant speedups without negatively impacting statistical accuracy. Similarly, \cite{wang2021fast} applied divide-and-conquer principles to optimize sparse Cox regression for high-dimensional survival data, showing substantial gains in computational efficiency. Task-based parallelism has also shown strong potential. \cite{hadjidoukas2015exploiting} used this approach in Bayesian uncertainty quantification, employing adaptive load balancing to schedule numerical differentiation and sampling tasks across multicore CPUs and GPUs. The result was a highly efficient, scalable framework for large-scale inference. In the context of metaheuristic optimization, \cite{santander2016asynchronous} implemented an asynchronous parallel model, allowing worker threads to proceed independently without synchronization barriers. This design reduced idle time and improved scalability, particularly for stochastic algorithms. For problems with strong data partitioning potential, \cite{yen2020solving} introduced a block-splitting method that reformulates fused penalty estimation into a separable structure, enabling efficient data parallelism. Numerous additional studies highlight the breadth of parallel design strategies applied in SC. These include a wide array of works based on divide-and-conquer strategies~\citep{bai2011divide, loh2011fast, kim2016parameter, sabnis2016divide, guhaniyogi2017divide, deng2017divide, xu2019distributed, minsker2019distributed, roy2019likelihood, su2020divide, sukkuea2022practical, hong2022divide, vyner2023swiss, chen2024simple}, task-based parallelism~\citep{canales2016large, ko2020diststat, abdulah2021accelerating, mondal2022parallel}, data parallelism~\citep{michalak2012developing, homrighausen2016nystrom}, and MapReduce-based approaches~\citep{liu2010mapreduce, mohammed2014applications, ferraro2019analyzing, mwamnyange2021big}. In summary, these works demonstrate the flexibility and effectiveness of parallel computing paradigms in addressing the computational demands of modern statistical problems, where they often achieve huge improvements in scalability and runtime without sacrificing accuracy.

%\begin{table}[h]
%    \centering
%    \renewcommand{\arraystretch}{1.3} % Adjust row height for better readability
%    \setlength{\tabcolsep}{8pt} % Adjust column spacing
%    \begin{tabular}{|l|l|}
%        \hline
%        \textbf{Design Pattern} & \textbf{References} \\ \hline
%        Divide and Conquer & \cite{guhaniyogi2017divide,su2020divide,sukkuea2022practical,deng2017divide,hong2022divide,kim2016parameter,chen2024simple,sabnis2016divide,loh2011fast,loh2011fast,vyner2023swiss,minsker2019distributed,bai2011divide,chen2024simple,xu2019distributed,roy2019likelihood,hong2022divide} \\ \hline
%        Task-based Parallelism & \cite{ko2020diststat,abdulah2021accelerating,canales2016large,mondal2022parallel,hadjidoukas2015exploiting} \\ \hline
 %       Data Parallelism & \citep{michalak2012developing,homrighausen2016nystrom,yen2020solving} \\ \hline
  %      Map-Reduce & ~\citep{liu2010mapreduce,ferraro2019analyzing,mohammed2014applications,mwamnyange2021big} \\ \hline
   % \end{tabular}
   % \caption{Parallel Design Patterns and Examples of Their Associated Statistical Literature}
   % \label{tab:design_patterns}
%\end{table}

\subsection{Data Management and Movement Optimization}

Efficient data management and movement are fundamental challenges in HPSC, where datasets grow to terabyte and petabyte scales. Traditional statistical workflows typically assume that data can be held in memory or processed sequentially, assumptions that break down in distributed, high-performance environments. At HPC scales, concerns such as data locality, partitioning strategy, and communication overhead become performance-critical in many cases, where the cost of moving data—whether across nodes or through memory hierarchies can exceed the benefits gained from parallel computation. Statistical tasks, such as subsampling, shuffling, or conditioning, often induce irregular memory access patterns that are difficult to optimize using conventional memory management techniques.

The complexity increases further when dealing with high-dimensional, heterogeneous data. Designing scalable data pipelines requires a smooth understanding of how to efficiently move and store data without compromising analytical accuracy. To meet these demands, researchers increasingly turn to out-of-core algorithms~\citep{yildirim2023iomax}, data compression techniques~\citep{kriemann2022high}, and communication-avoiding methods~\citep{shohdy2016fault}. Some strategies, such as locality-aware partitioning {\color {black} (i.e., carefully distributing data across processing units to maximize data locality)} and asynchronous data prefetching {\color {black} (i.e., loading data into memory before they are needed for computation)}, aim to reduce latency and improve throughput, though often at the cost of added complexity or overhead. Ultimately, constructing data pipelines that align with the architecture’s computational and memory hierarchy is essential for unlocking the full potential of HPSC. As such, data movement is not merely a systems concern; it is a central design constraint in SC at scale that can heavily impact performance.

%%%%%%%%%%%STOP
\subsection{Numerical Stability and Precision Issues}

Numerical stability is a critical challenge in SC, and its importance grows with the scale and complexity of modern HPC workloads~\citep{altman2004numerical}. Large-scale computations amplify the impact of rounding errors, cancellation, and other numerical artifacts that may be negligible at smaller scales. Statistical methods such as Bayesian inference, maximum likelihood, and covariance matrix inversion are sensitive to small errors in input data or intermediate results. {\color {black} When computations are distributed across thousands of processing units, tiny floating-point rounding errors from individual operations can accumulate, sometimes leading to noticeable differences in large-scale results. While this is a general concern, today’s advancements in low-precision computing, driven by GPUs and specialized accelerators, introduce an additional layer of complexity. Lower-precision formats, such as FP16 and BFLOAT16, offer significant improvements in performance and energy efficiency; however, these gains often come at the expense of reduced numerical accuracy.} This accuracy degradation is especially problematic for iterative solvers, gradient-based optimization, and matrix decompositions, which form the backbone of many statistical algorithms~\citep{abdulah2021accelerating, salvana2024mpcr}. Balancing speed and robustness in this setting requires a careful redesign of the algorithm. One promising mitigation strategy is the use of stochastic rounding, which reduces systematic bias introduced by low-precision arithmetic and improves the stability of iterative computations at scale~\citep{croci2022stochastic}. However, the challenges go beyond precision. Reproducibility becomes increasingly a problem in parallel environments, where thread scheduling, non-deterministic reductions, and hardware-level optimizations can lead to variations in results. In statistical applications, especially those with downstream scientific or policy implications, ensuring numerical and statistical reproducibility is not optional; it is a foundational requirement. Addressing these concerns requires both algorithmic safeguards and systems-level support for deterministic, high-precision computing when needed.

\subsection{Software Design for Portability and Reproducibility}
Designing statistical software for HPC requires a careful balance between performance, portability, and reproducibility. Portability means that code can run efficiently across diverse architectures, including CPUs, GPUs, and emerging accelerators, without requiring extensive rewrites. Achieving this typically involves adopting abstraction layers and performance-portable programming models, such as Kokkos~\citep{sunderland2016overview} and RAJA~\citep{hornung2014raja}, which decouple hardware-specific details from algorithmic logic. In addition to performance-portable programming models, software portability can be further enhanced through the use of containerization. {\color {black} Containerization is a lightweight virtualization technique that packages an application together with its dependencies, libraries, and runtime into a single portable unit.} Containerization tools, such as Docker and Singularity~\citep{schulz2016use, abdulah2024portability}, also play a key role, enabling consistent execution environments across platforms while simplifying deployment and dependency management.

Moreover, reproducibility is essential, particularly in statistical applications where numerical results directly inform scientific conclusions or policy decisions. Ensuring reproducibility requires careful control over software environments, versioning, data provenance, and the generation of random numbers. This often involves workflow orchestration systems, standardized data formats, and comprehensive logging of configuration states. When these practices are integrated early in the development lifecycle, teams can build HPC-ready statistical software that is not only high-performing but also robust, maintainable, and verifiable. Embedding portability and reproducibility as core design goals is fundamental to advancing reliable HPSC.

\subsection{Implementation Challenges on Heterogeneous Architectures}
Deploying statistical algorithms on heterogeneous HPC architectures, including CPUs, GPUs, and specialized accelerators, also presents numerous practical implementation challenges. These systems demand explicit control over data movement across complex memory hierarchies, careful load balancing, and hardware-specific performance tuning. Statistical workloads are often characterized by irregular computation, conditional branching, and fine-grained data dependencies, which do not map cleanly onto massively parallel devices, such as GPUs. As a result, porting statistical algorithms typically requires rethinking about the core data structures, optimizing memory access patterns, and integrating parallel libraries such as cuBLAS {\color{black} (GPU-accelerated linear algebra)}, Thrust {\color{black} (parallel algorithms and data structures)}, or oneAPI {\color{black} (cross-architecture portability)} to extract performance. Portability adds further complexity. Achieving efficient execution across vendor-specific platforms, such as NVIDIA's CUDA and AMD's ROCm, requires either developing separate backends or using abstraction layers that can introduce overhead or limit optimization. Addressing these issues demands a modular software architecture that cleanly separates hardware-agnostic logic from performance-critical kernels, enabling both scalability and maintainability across diverse systems. Language-level limitations can further impede adoption. For example, R, the widely used programming language in the SC community, does not directly support GPU computing or parallel execution on heterogeneous systems. This gap underscores the need for broader community investment in extending R's capabilities for high-performance environments and heterogeneous distributed systems. Efforts, such as the \textit{MPCR} package for multi- and mixed-precision computing~\citep{salvana2024mpcr}, the \textit{future} package in R, which offers a unified and extensible framework for concurrent and parallel programming with a conceptually simple and expressive syntax~\citep{bengtsson2020unifying}, and the integration of R with the COMPSs runtime system for distributed execution via the \textit{RCOMPSs} package~\citep{zhang2025rcompss}, represent important steps forward. However, these tools are still in their early stages and require further development, optimization, and broader community adoption to fully enable scalable statistical computing on modern, heterogeneous HPC platforms.

%%%%%%%%%%%%%%%%%%%%%%%%%%%%%%%%%%%%%%%%%%%%%%%%%%%%%%%%%%%%%%%%%%
\section{HPC-Driven Opportunities in Statistical Computing}
\label{sec:opportunities}
%HPC offers a transformative pathway for scaling up the speed and the accuracy of statistical computing. By leveraging the power of distributed architectures, accelerators, and high-throughput pipelines, HPC enables the design and execution of statistical methods that were out of reach, whether due to data volume, algorithmic complexity, or computational cost. These capabilities open the door to scalable Bayesian inference, high-dimensional simulation, real-time modeling, and robust uncertainty quantification on massive datasets.
In this section, we examine how HPC extends the boundaries of what is possible in SC. It highlights key opportunities for accelerating inference, enhancing model fidelity, and reducing reliance on heuristics or coarse approximations. From enabling finer-grained analysis to supporting complex, multiscale models, HPC is not only an enabler of speed, it is a motivation for reimagining the scale and scope of statistics itself.

\subsection{Parallel Statistical Algorithms}

Parallel statistical algorithms are foundational to advancing computational workflows in modern, data-intensive applications. Many core methods, such as Monte Carlo simulations, bootstrap resampling, expectation-maximization (EM), and MCMC, naturally lend themselves to parallel execution due to their iterative structure and inherent concurrency~\citep{suchard2010understanding, bottou2018optimization}. Distributing these computations across multicore CPUs, GPUs, or hybrid architectures dramatically reduces runtime and unlocks scalability for larger datasets and increasingly complex models. This capability enables faster convergence, richer model exploration, and real-time inference across diverse domains—including Bayesian inference, spatial statistics, biostatistics, and machine learning. Realizing these gains, however, demands not just access to powerful hardware but also algorithmic innovation and careful alignment between statistical fidelity and implementation strategy.

Designing efficient parallel algorithms requires thoughtful orchestration of memory, communication, and computation. Modern approaches often blend shared- and distributed-memory paradigms, optimizing for cache locality, interconnect bandwidth, and memory hierarchy~\citep{rauber2013parallel}. Load balancing, whether through static partitioning or dynamic strategies such as work stealing {\color {black} (i.e., idle processors ``steal" tasks from the queues of busy processors, improving overall load balance)}, is essential to ensure high utilization, particularly on heterogeneous systems~\citep{blumofe1999scheduling}. As systems grow more complex, adaptive scheduling and intelligent resource allocation become increasingly important. Hardware architecture significantly shapes performance. Optimizing for multicore CPUs, GPUs, and emerging accelerators entails understanding their respective execution models, memory layouts, and compute hierarchies~\citep{abdelfattah2016high, ashari2014fast}. Heterogeneous systems exacerbate these challenges, necessitating hybrid execution strategies that dynamically coordinate across different device types and workloads.

A robust and evolving software ecosystem supports parallel statistical computing. Low-level models such as MPI, OpenMP, and CUDA/OpenCL provide fine-grained control over distributed, shared, and GPU-based execution, while high-level frameworks like \texttt{Dask}, \texttt{future}, and parallel extensions of R and Python help abstract complexity~\citep{rocklin2015dask}. However, reproducibility and numerical stability remain nontrivial. Parallel execution introduces sensitivity to floating-point accumulation, rounding errors, and nondeterministic execution orders~\citep {higham2002accuracy}. 

Achieving peak performance requires minimizing data movement and maximizing computational throughput. The Roofline model~\citep{williams2009roofline} offers a guiding framework for this trade-off, while strong and weak scaling analyses help evaluate algorithmic efficiency across architectures and problem sizes. Recent advances explore communication-avoiding algorithms \citep{xu2024numerically,katagiri2024communication}, adaptive parallelism responsive to resource availability, and domain-specific abstractions that simplify implementation without impacting performance. As HPC systems continue to scale, resilience becomes a key concern. Techniques such as checkpoint-restart and algorithm-based fault tolerance are increasingly becoming standard~\citep{shohdy2016fault}. Ultimately, the evolution of parallel statistical algorithms lies at the intersection of hardware innovation, software engineering, and statistical theory. As problem sizes grow and models become increasingly complex, scalable parallelism will remain a defining pillar of HPSC.

\subsection{Parallel Dense and Sparse Linear Algebra Libraries}

Parallel dense and sparse linear algebra libraries are essential enablers of scalable statistical computing in the HPC landscape. Many core statistical algorithms, ranging from regression and principal component analysis to high-dimensional model fitting, rely heavily on matrix operations. These libraries provide highly optimized, parallel implementations of operations such as matrix factorizations (LU, QR, Cholesky) and eigenvalue decompositions, tailored for multicore and manycore architectures. For shared-memory systems, libraries such as \textit{OpenBLAS}~\citep{openblas}, \textit{Intel MKL}~\citep{mkl}, \textit{AMD BLIS}~\citep{blis}, and \textit{NVIDIA cuBLAS}~\citep{cublas} harness thread-level parallelism and vectorized instructions. For distributed and GPU-accelerated environments, tools like \textit{ScaLAPACK}~\citep{scalapack}, \textit{Elemental}~\citep{elemental}, \textit{DPLASMA}~\citep{dplasma}, and \textit{MAGMA}~\citep{magma} enable large-scale matrix computations by distributing workloads across nodes or GPU devices. These libraries form the computational backbone of many statistical workflows, delivering substantial gains in speed, memory efficiency, and scalability.

Sparse linear algebra libraries are equally vital, particularly for applications involving high-dimensional, yet sparse, data structures. These tools are designed to minimize memory use and accelerate computation by exploiting sparsity patterns in matrices. Libraries such as \textit{SuiteSparse}~\citep{davis2019algorithm} offer efficient routines for sparse matrix factorization and graph-based computations. \textit{Intel MKL} includes optimized sparse BLAS and solvers for Intel hardware, while \textit{cuSPARSE} delivers high-performance sparse operations on NVIDIA GPUs. \textit{PETSc} (Portable, Extensible Toolkit for Scientific Computation) provides scalable sparse solvers, preconditioners, and optimization routines, and is widely adopted in both scientific computing and machine learning applications~\citep{balay2019petsc}. Likewise, \textit{Trilinos}~\citep{heroux2005overview} offers a flexible framework for building scalable, domain-specific solvers in HPC environments. Together, these libraries support statistical methods such as sparse regression, large-scale generalized linear models, and spatial modeling—delivering high performance while maintaining numerical stability and accuracy.

A growing body of work demonstrates how these libraries are being integrated into statistical software ecosystems. For example, \cite{yoginath2005rscalapack} introduced \textit{RScaLAPACK}, which connects the R environment with ScaLAPACK to enable distributed linear algebra without requiring users to manage parallelism explicitly. Implemented in C and Fortran with MPI, it enables scalable matrix operations such as eigenvalue decompositions and SVD while preserving R’s familiar syntax. \textit{RScaLAPACK} has been adopted in studies such as \cite{elgamal2015analysis} for large-scale statistical analysis and SVD computation. %Similarly, the \textit{ExaGeoStat} framework~\citep{abdulah2018exageostat} was developed to support large-scale spatial modeling using dense linear algebra kernels. It leverages optimized BLAS/LAPACK routines across heterogeneous architectures and has since evolved to support Tile Low-Rank (TLR) approximations~\citep{abdulah2018parallel}, mixed-precision execution~\citep{abdulah2019geostatistical,cao2022reshaping,cao2023reducing}, and non-Gaussian models~\citep{mondal2023tile}. The framework has also been wrapped in R~\citep{abdulah2023large}, expanding accessibility for applied users. 
Another important contribution is the \textit{bigGP} package~\citep{paciorek2015parallelizing}, which introduces scalable GP modeling to R by leveraging ScaLAPACK, BLAS, and MPI. By distributing Cholesky decompositions and covariance matrix operations across nodes, \textit{bigGP} enables efficient kriging, likelihood optimization, and high-dimensional GP regression that would otherwise be computationally prohibitive.

Sparse matrix libraries have also been successfully integrated into statistical applications. For instance, \cite{castrillon2016multi} presents \textit{SuiteSparse} to accelerate large-scale sparse regression and covariance tapering methods, reducing runtime while preserving analytical fidelity. In biostatistics, Raim et al. (2013) employed the Portable, Extensible Toolkit for Scientific Computation (\textit{PETSc})~\citep{balay2019petsc} and its toolkit for advanced optimization (\textit{TAO})~\citep{benson2003tao} to parallelize maximum likelihood estimation for the random-clumped multinomial model. By distributing the optimization workload across an HPC cluster, their method achieved scalable, efficient inference on large datasets.

Altogether, parallel dense and sparse linear algebra libraries represent an indispensable infrastructure for HPSC. They provide the foundation for building scalable algorithms that can fully exploit modern hardware, bridging the gap between statistical sophistication and computational efficiency.

\subsection{HPC Tools for Big Data and Streaming Statistical Analysis}

Statistical applications increasingly operate at data scales that strain or exceed the limits of traditional computing. Domains such as spatial statistics have seen explosive growth in data volume due to advances in satellite sensing, remote imaging, and IoT technologies~\citep{brunsdon2020big}. Similar trends are evident in biostatistics, where genomics, medical imaging, and electronic health records (EHR) systems now generate petabyte-scale datasets~\citep{li2024exploring}. The high volume and velocity of modern datasets necessitate high-performance solutions. Conventional systems, constrained by memory and compute limits, often force practitioners to rely on approximations, sampling, or data reduction techniques. While these strategies are helpful, they can also limit the accuracy. By contrast, HPC offers a path toward exact, scalable analysis, delivering both speed and accuracy for high-dimensional problems.  Traditional models, designed for small, structured data, struggle with the complexity and scale of these sources. For instance, Genomic data often features millions of variables (e.g., gene expressions, mutations) across relatively few samples, creating challenging high-dimensional, low-sample-size (HDLSS) regimes~\citep{he2017big}. HPC enables practitioners to bypass many of these limitations, allowing scalable inference, full-resolution modeling, and faster iteration cycles across big data landscapes.

Distributed computing frameworks are central to HPC-enabled big data analytics. By distributing data and computation across nodes, these systems provide the necessary memory and throughput to manage massive datasets. However, many statisticians lack training in parallel programming or system-level tuning, creating a barrier to entry. Bridging this gap requires high-level abstractions that expose HPC performance without demanding low-level expertise. Several frameworks have emerged to meet this need. Apache Spark offers a widely adopted, in-memory data processing engine for large-scale analytics, with native support for machine learning and graph computation~\citep{salloum2016big}. Dask brings parallelism to Python workflows, scaling familiar libraries such as NumPy, pandas, and scikit, learn to work in distributed settings~\citep{rocklin2015dask}. Ray provides a flexible, high-performance execution framework built for parallel, AI-centric workloads, and has found growing use in statistical modeling~\citep{moritz2018ray}.

Despite their potential, the adoption of these tools within the statistical community remains limited, largely due to a lack of integration with traditional statistical environments and workflows. To close this gap, domain-specific frameworks have emerged that combine statistical rigor with HPC efficiency. For example, \textit{XGBoost}~\citep{chen2016xgboost} and \textit{LightGBM}~\citep{ke2017lightgbm} offer distributed, high-speed implementations of gradient-boosted trees, optimized for performance on high-dimensional datasets. These tools are increasingly used in applied statistics, epidemiology, and econometrics, providing scalable alternatives to conventional regression and classification models. As statistical workflows evolve, the continued development and adoption of HPC-compatible tools will be essential to fully leverage distributed computing resources in big data settings.

HPC also opens the door to real-time statistical analysis—an emerging priority in domains where decisions must be made with minimal latency. Traditional batch-processing methods are too slow for applications such as patient monitoring, financial risk management, or emergency response. Stream processing frameworks, coupled with HPC backends, enable rapid model updates, adaptive inference, and low-latency decision-making. % \textcolor{black}{In environmental statistics, real-time Kriging and Bayesian spatial modeling have been applied to high-frequency pollution data from IoT sensors in smart cities~\citep{kaivonen2020real}, allowing near-instantaneous detection of air quality anomalies.} 
%In the field of Digital Earth, a real-time ingestion and processing approach in Geospatial Data Cubes \citep[RTGDC,][]{LiuRTGDC} has been developed using a distributed streaming computing framework to enhance real-time capabilities. 
%In emergency response, statistical tools guide rapid resource allocation during disasters such as wildfires or earthquakes, \textcolor{black}{relying on real-time spatial statistics~\citep{kwan2005emergency}}. 
Biostatistics presents further applications: continuous patient monitoring can detect subtle physiological changes by analyzing high-volume sensor data from ICU devices~\citep{etli2024future}, while real-time surveillance models track infectious disease outbreaks using hospital records, mobile apps, and public health databases~\citep{desai2019real}.

Together, these developments underscore the growing importance of HPC in both batch and streaming statistical analytics. By enabling large-scale, low-latency computation, HPC is reshaping the boundaries of what is computationally feasible, expanding the reach, speed, and impact of modern statistical science.

\subsection{Energy-Efficient Statistical Computing}

As HPC systems scale in processor count, memory capacity, and accelerator usage, energy consumption has become an increasingly pressing concern, both environmentally and economically. Large-scale statistical applications, including simulation-based inference, bootstrapping, and iterative optimization, often incur high computational costs that translate directly into significant energy usage. Designing energy-efficient statistical workflows is therefore critical, particularly as sustainability becomes a central consideration in scientific computing. One promising approach is mixed-precision computation, where select operations are executed in lower-precision formats (e.g., FP16 or bfloat16), reducing power consumption without compromising statistical accuracy. This strategy is especially effective on GPUs and AI accelerators optimized for low-precision arithmetic~\citep{dongarra2024hardware}. Mixed-precision methods have proven successful in accelerating iterative algorithms such as MLE, achieving faster convergence and reduced energy footprints~\citep{cao2023reducing}.

Beyond precision tuning, algorithmic approximation techniques offer further gains in energy efficiency. Randomized numerical linear algebra methods—such as sketching, random projections, and low-rank matrix approximations—can significantly reduce computational load while maintaining acceptable accuracy~\citep{baumann2024energy, abdulah2018parallel}. Likewise, methods like approximate Bayesian computation (ABC)~\citep{csillery2010approximate} and variational inference serve as scalable alternatives to traditional MCMC, enabling faster inference under limited resource budgets. These approximations are especially valuable in high-dimensional or streaming settings, where evaluating the whole model is often infeasible. As discussed throughout this review, such strategies are already making statistical computing more tractable across domains ranging from spatial modeling to genomics.

Hardware-aware optimization further amplifies these benefits. Adapting statistical algorithms to energy-efficient architectures—such as ARM processors~\citep{ou2012energy} or vectorized instruction sets~\citep{jakobs2016reducing}—can yield substantial savings without sacrificing performance. Minimizing data movement is another key priority, as communication across memory hierarchies or networked systems often accounts for a significant portion of energy costs. Techniques such as communication-avoiding algorithms and asynchronous execution can mitigate these bottlenecks, enabling more sustainable computation at scale.

As demand grows for large-scale statistical analysis in fields like epidemiology, climate science, and biomedicine, incorporating energy efficiency into both algorithm and system design is no longer optional. It is a necessary step toward building a statistical computing infrastructure that is not only fast and accurate but also environmentally and economically sustainable.

%%%%%%%%
%R2: Distinguish the near-term, low-risk achievable future directions from the long-term speculative directions that involve significantly higher uncertainty. 
\section{Future Directions}
\label{sec:future}
%HPSC is driven by the dual forces of rapidly expanding data volumes and rapid advancements in computing hardware and software.
As scientific, engineering, and industrial applications generate ever more complex and massive datasets, the statistical community faces both pressing challenges and unprecedented opportunities: to scale existing methods, design new algorithms, and reimagine statistical inference at the extremes of scale. Meeting these demands will require more than technical progress alone; it calls for deep, sustained collaboration between statisticians, computer scientists, and HPC practitioners. It also demands the development of new tools, standards, and methodologies that bridge disciplinary divides and make HPC accessible, robust, and reproducible for statistical applications. This section highlights several new directions shaping the future of HPSC, including the integration of specialized hardware, the rise of federated statistical computing, the support for standardization and interoperability, and the creation of statistical methods for modern high-performance environments. {\color {black} While many opportunities exist for advancing HPSC, it is helpful to distinguish between directions that are achievable in the near term with relatively low risk and those that are longer-term and more speculative, involving higher uncertainty but potentially transformative impact.}

\subsection{Statistical Computing on Specialized Hardware (GPU, TPU, IPU, DPUs)}

Most existing R libraries and statistical software are optimized for CPU execution and offer limited support for GPU acceleration, typically through low-level C/C++ bindings. As datasets rapidly grow and models become more complex, this CPU-centric approach is increasingly insufficient. Tasks such as large-scale simulation, Bayesian inference, and high-dimensional modeling often exceed the capabilities of a single-node CPU, resulting in significant performance bottlenecks. Modern accelerator hardware, initially designed for machine learning, offers massive parallelism and high memory bandwidth that can dramatically speed up statistical workloads. GPUs, in particular, enable large-scale matrix operations and sampling routines to be executed orders of magnitude faster. By employing low-precision formats such as FP16 or BF16 where acceptable, and reserving FP32 or FP64 for numerically sensitive operations, modern platforms can cut memory usage, boost throughput, and reduce energy consumption~\citep{ltaief2023steering, cao2023reducing}. However, statistical computing often requires high numerical accuracy, especially for tasks such as variance estimation, likelihood evaluation, and posterior sampling. Thus, future work must focus on adaptive mixed-precision strategies that dynamically adjust precision to preserve stability and avoid bias in intermediate computations~\citep{abdulah2024boosting}.

The shift toward specialized hardware also opens the door to domain-specific accelerators. GPU tensor cores~\citep{markidis2018nvidia}, Intelligence Processing Units \citep[IPUs,][]{louw2021using}, and Data Processing Units \citep[DPUs,][]{barsellotti2022introducing} offer compelling advantages for different classes of statistical computation. IPUs are well-suited for fine-grained parallelism in matrix-intensive tasks such as GPs and hierarchical Bayesian models. DPUs, on the other hand, can offload data movement and preprocessing operations—like shuffling, sorting, and filtering—freeing up the main processor for model training or inference. Looking further ahead, quantum computing presents a radically different paradigm. While still in its infancy, quantum accelerators have the potential to offer exponential speedups for certain statistical tasks~\citep{yazdi2024application}. Early exploration into quantum algorithms, such as quantum Monte Carlo methods, variational quantum eigensolvers~\citep{tilly2022variational}, and quantum-enhanced optimization—could unlock new approaches to Bayesian inference, MCMC, and high-dimensional clustering. Major challenges remain, including hardware instability, high error rates, and the need for hybrid quantum–classical workflows. However, investing in quantum circuit design, error correction techniques, and algorithmic robustness could now position the statistical computing community to fully leverage the benefits as quantum technologies mature. {\color{black}  This is an achievable future direction for statisticians that requires understanding modern hardware architectures and optimizing existing statistical algorithms to fit these architectures well.}

%%%%%%%%%%%%%
\subsection{Federated Statistical Computing and Privacy-Preserving Inference}

Initially developed in the machine learning community~\citep{mcmahan2017communication}, federated learning enables decentralized model training across multiple data sources without requiring raw data to be centrally aggregated. Building on this foundation, the emerging paradigm of Federated Statistical Computing (FSC) extends these principles to traditional statistical modeling. FSC provides a framework for performing inference across distributed datasets, including health records, financial transactions, and climate measurements, while respecting privacy constraints and regulatory boundaries~\citep{jordan2019communication}. Instead of pooling raw data, federated statistical workflows compute local sufficient statistics, parameter estimates, or posterior approximations at each institution and securely aggregate them to form a global model. This preserves privacy while enabling reproducible, interpretable, and statistically valid inference. Active areas of research include the integration of privacy-preserving technologies such as differential privacy~\citep{dwork2009differential} and homomorphic encryption~\citep{lu2016using}, which provide strong formal guarantees for secure computation. Recent advances in federated Bayesian modeling~\citep{zhang2022personalized} and federated GPs~\citep{yue2024federated} demonstrate how complex probabilistic models can be trained without exposing sensitive data or intermediate computations.

\begin{figure*}
\centering
    \includegraphics[width=0.4\textwidth]{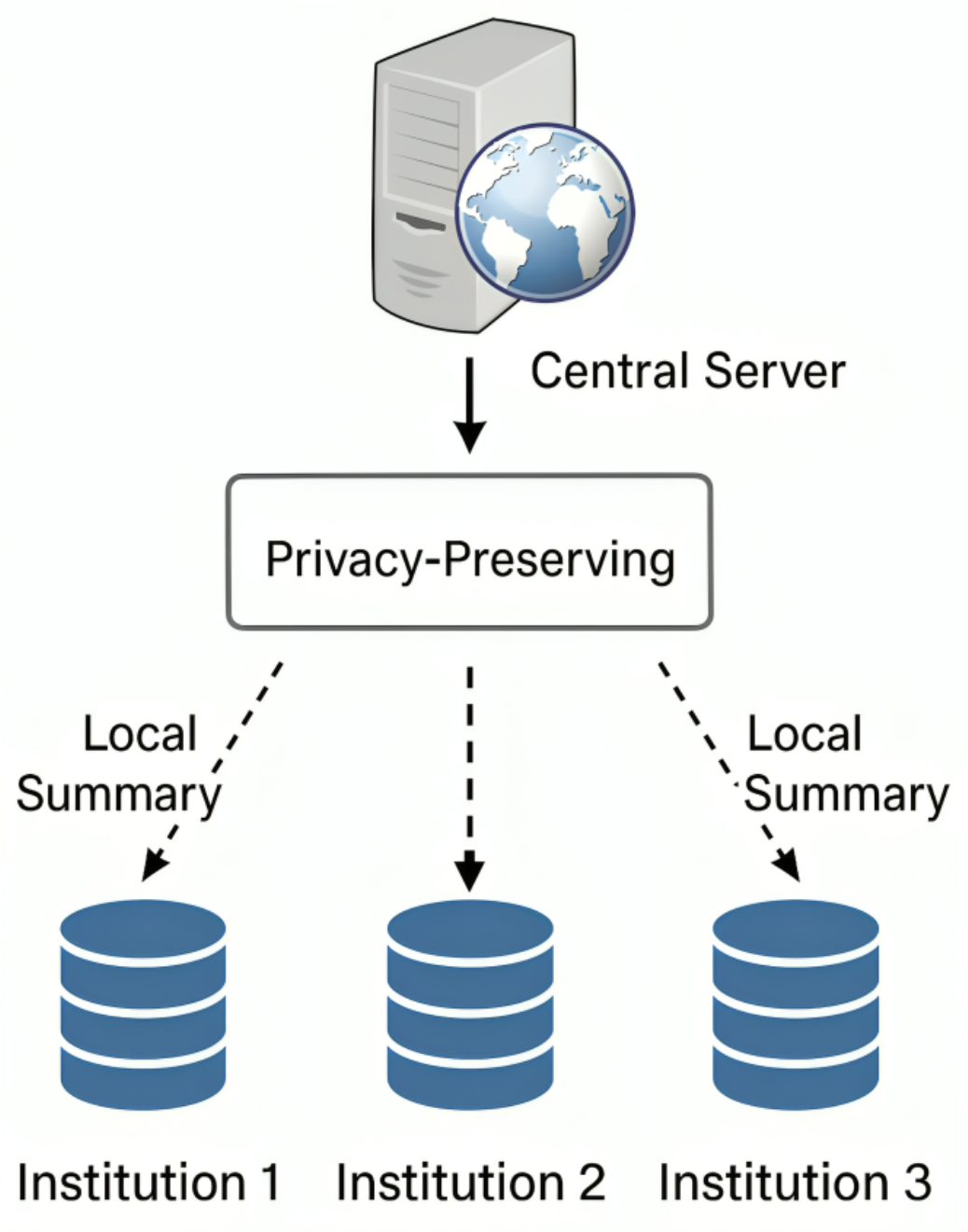} % Change filename
    \captionof{figure}{Federated Statistical Computing: Institutions share local statistical summaries with a central server to enable privacy-preserving model aggregation.}
    \label{fig:fsc}
\end{figure*}

However, FSC introduces challenges distinct from those in federated learning. Statistical inference demands accurate uncertainty quantification, posterior consistency, and asymptotic guarantees—all of which must be revisited in settings characterized by communication bottlenecks, heterogeneous data distributions, and asynchronous updates across institutions. Novel algorithms, such as federated stochastic gradient Langevin dynamics~\citep{el2021federated} and hierarchical model partitioning~\citep{marion2018hierarchical}, aim to minimize communication overhead while maintaining statistical integrity. Applications of FSC are rapidly expanding into domains such as survival analysis, spatial modeling, and causal inference—especially in contexts where collaborative analysis is essential but centralized data collection is infeasible, as seen in healthcare, finance, and environmental science. As FSC evolves, it is poised to become a cornerstone of scalable, privacy-aware statistical computing. Future directions include developing federated analogs to classical statistical procedures, formalizing uncertainty-aware aggregation protocols, and integrating FSC into HPC workflows. In doing so, FSC closes a crucial gap between statistical methodology and the decentralized, privacy-constrained reality of modern data science. Figure~\ref{fig:fsc} illustrates a typical FSC architecture, where multiple institutions compute and securely transmit local summaries—such as histograms, regression coefficients, or scatter plots—to a central server for privacy-preserving model integration. {\color{black}  FSC is an achievable, low-risk direction. Recent studies are tackling privacy challenges, and the mature ML literature on federated learning offers clear guidelines for researchers.}

%%%%%%%%%%%%%%%%%%%%%%%%%%%%%%%%%%%%
\subsection{Standardization and Interoperability}

As HPSC continues to expand, the need for standardization and interoperability has become increasingly critical. The current statistical computing ecosystem remains highly fragmented, composed of specialized libraries, programming languages, and toolchains optimized for narrow domains, specific hardware architectures, or bespoke workflows. This heterogeneity creates substantial barriers to reproducibility, portability, and collaborative development. Looking forward, one of the most pressing priorities is the development of standardized interfaces for distributed statistical computing—interfaces that enable algorithms to operate seamlessly across platforms, from local machines to large-scale HPC clusters. Such standards would define consistent protocols for data exchange, task coordination, and metadata management, allowing statistical workflows to scale without the need for extensive code rewrites. Advances in distributed computing frameworks—from MPI-based systems to modern, task-parallel runtimes—are already laying the foundation for this interoperability.

Equally important is the development of portable statistical software that performs reliably across a range of HPC architectures, including multicore CPUs, GPUs, and emerging accelerators, each with its own programming model and performance characteristics. High-level abstraction layers, such as Kokkos and SYCL, allow developers to write performance-portable code. Meanwhile, containerization tools such as Docker and Singularity help encapsulate statistical environments, ensuring reproducibility and simplifying deployment across diverse systems. Integrating these technologies into mainstream statistical tools is crucial for their widespread adoption. For example, while R remains a cornerstone of statistical practice due to its accessible syntax and rich package ecosystem, it lacks native support for distributed or GPU computing. Bridging this gap often requires wrapping low-level HPC libraries using tools such as Rcpp or exposing high-performance backends through user-friendly interfaces. Similarly, efforts to integrate machine learning frameworks, such as TensorFlow, with R and Python demonstrate how statistical computing can benefit from shared computational infrastructure. These integrations not only accelerate performance but also enable new hybrid workflows that combine traditional statistical modeling with scalable, learning-based methods. Standardization and interoperability are thus not peripheral concerns—they are foundational to making HPSC broadly usable, maintainable, and impactful in real-world applications.

{\color{black} In the near term, low-risk progress in this direction lies in standardizing distributed interfaces and adopting performance-portable layers and containers. Longer-term, higher-uncertainty goals include a cross-language intermediate representation for statistical kernels, automated cross-platform autotuning, and end-to-end standards across HPC–cloud–edge and federated settings, all of which demand substantial community coordination and new tooling.}

\subsection{Novel Statistical Methods}

The future of HPSC is not only about retrofitting existing algorithms to run on faster hardware; it increasingly demands the creation of new statistical methods explicitly designed for parallel, distributed, and heterogeneous computing environments. Many classical techniques were built for sequential execution, with tight interdependencies between computational steps. As data volumes grow exponentially and hardware becomes more parallel, these assumptions no longer hold. This shift presents both a challenge and an opportunity: to rethink core statistical approaches from the ground up for modern computing architectures.

One promising direction involves methods that are inherently parallel, rather than sequential algorithms awkwardly adapted to parallel settings. In scalable Bayesian inference, for instance, techniques like Consensus Monte Carlo \citep {scott2017comparing} compute independent posterior samples across data subsets and combine them post hoc, enabling efficient inference across distributed datasets. Similarly, divide-and-conquer strategies for likelihood estimation and optimization \citep{zhang2013divide} partition computations into smaller, parallelizable subtasks, yielding dramatic runtime reductions for large-scale problems.

Approximation methods also play an important role. Methods such as variational inference, randomized sketching, and low-rank matrix approximations help to strike a balance between computational efficiency and statistical accuracy. Algorithms such as randomized singular value decomposition~\citep{halko2011finding} and sketching-based regression~\citep{woodruff2014sketching} make it feasible to analyze massive datasets within memory and time constraints. Although approximations can introduce bias or affect uncertainty quantification, they are often essential to enable inference at scale. The growing convergence of statistical modeling and machine learning within HPC contexts is also driving hybrid methodologies that fuse the interpretability of statistical inference with the representational power of deep learning. Examples include deep Bayesian neural networks, probabilistic graphical models on GPUs, and statistical layers embedded within neural architectures—tools increasingly deployed in scientific domains where uncertainty and explainability are paramount.

At scale, communication costs become a significant bottleneck. Efficient statistical computing requires novel methods to manage this overhead, particularly for iterative or gradient-based algorithms. Communication-avoiding techniques, asynchronous updates, and decentralized learning strategies are actively being developed to mitigate these constraints~\citep{ballard2014communication}. In tandem, hierarchical modeling frameworks that map naturally onto HPC hardware hierarchies—nodes, sockets, cores—can minimize costly inter-node data movement.

Emerging hardware platforms, such as GPUs, TPUs, and IPUs, offer new opportunities for designing statistical algorithms. Tools such as \textit{ExaGeoStat}\citep{abdulah2018exageostat} and \textit{pbdR}\citep{Ostrouchov2020} show how high-performance linear algebra and GPU acceleration can power scalable spatial and multivariate inference. Future research must continue to develop statistical kernels optimized for tensor cores, mixed-precision arithmetic, and hardware-specific instructions to ensure performance and accuracy on next-generation systems.

Finally, debugging and validation of such methods become more complex as they scale. Ensuring correctness, numerical stability, and reproducibility across distributed architectures remains a nontrivial task. Building robust diagnostic tools, test suites, and reproducible benchmarking frameworks will be essential to gaining user trust and accelerating adoption. Continued co-design between statistical methodology and HPC will shape the next generation of tools and theory, keeping statistical science not only relevant but central to the future of large-scale, data-driven discovery.

\section{A Roadmap for the HPSC Community}
\label{sec:roadmap}
The convergence of HPC and statistical science holds transformative potential for modern data analysis. However, realizing this potential will require more than technical innovation; it demands the cultivation of a dedicated, interdisciplinary community. Building such a community means bridging cultural and technical divides, fostering sustained collaboration, and training a new generation of researchers with deep expertise in both advanced statistical modeling and HPC technologies. These researchers will play a crucial role in developing novel methods that can handle the scale, complexity, and computational demands of contemporary scientific and industrial data. As data-intensive applications across genomics, climate modeling, remote sensing, finance, and public health continue to outpace the capabilities of traditional statistical workflows, the need for scalable, high-performance statistical inference has never been more urgent. Sophisticated models, such as high-dimensional hierarchical Bayes, and real-time applications, including anomaly detection and epidemiological forecasting, increasingly rely on the computational power and architectural diversity that HPC environments provide.

Encouragingly, progress is already underway. Efforts such as the \textit{pbdR} suite~\citep{hasan2019scalable}, \textit{ExaGeoStat}~\citep{abdulah2018exageostat}, and other high-performance libraries demonstrate the viability of scalable statistical software that harnesses distributed-memory systems and accelerators. Interdisciplinary sessions at conferences such as Supercomputing (SC), Joint Statistical Meetings (JSM), and SIAM are also fostering dialogue between statisticians and computational scientists. However, these efforts remain fragmented—often limited to isolated projects without shared standards, cohesive tooling, or broad adoption. For HPSC to mature into a unified ecosystem, the community must move from isolated success stories to an integrated framework that supports sustainable growth and wide applicability.

HPC offers unparalleled capabilities, including high-throughput computation, access to GPUs, TPUs, and IPUs, as well as powerful tools for profiling and optimizing performance. Statistical science, in turn, offers essential contributions to uncertainty quantification, dimensionality reduction, reproducibility, and interpretability—core elements often lacking in traditional HPC applications. Yet integration is still hindered by several structural barriers. HPC resources are often inaccessible to statisticians due to administrative overhead, institutional gatekeeping, and unfamiliar software stacks. Most statisticians work in interactive, high-level languages such as R or Python, while HPC workflows tend to rely on batch processing, lower-level languages, and rigid scheduling systems. Additionally, a few individuals possess deep expertise in both domains, creating knowledge silos that slow collaboration and limit methodological innovation. {\color {black} A key step toward overcoming these barriers is the formation of a dedicated community around HPSC. To this end, the recently launched webpage \footnote{\url{https://hpsc4science.org}} serves as a central hub for building this community. The platform aims to connect statisticians, data scientists, and HPC scientists by organizing workshops, short courses, collaborative projects, and online discussions. This should provide visibility for emerging tools, encourage mentorship across career stages, and promote interdisciplinary engagement. By establishing a common identity and shared resources, this community-driven effort will accelerate the maturation of HPSC from scattered initiatives into a cohesive field with global impact.
}

Herein, we also propose a roadmap to address these challenges centered around five pillars: dialogue, education, software, reproducibility, and funding.

\begin{enumerate}
    \item \textit{Dialogue}: Establish sustained communication channels between HPC and statistical communities through interdisciplinary panels, workshops, and joint research initiatives. These forums can surface shared challenges, accelerate collaboration, and catalyze co-designed solutions.

    \item \textit{Education}: Develop interdisciplinary training programs that equip researchers with foundational HPC knowledge along with statistical theory. Curriculum modules, summer schools, and online courses can prepare the next generation to work fluently across both domains.

    \item \textit{Software and Standards}: Promote open-source tools and standardized APIs that seamlessly integrate statistical methods into HPC workflows. Create shared repositories of reproducible case studies, scalable datasets, and benchmark metrics to guide adoption and experimentation.

    \item \textit{Reproducibility and Accessibility}: Encourage containerized deployments, high-level interfaces, and performance-portable frameworks that reduce the barrier to entry for statisticians. HPC researchers should develop interactive tools, robust documentation, and reusable modules that align with statistical workflows and prioritize user experience.

    \item \textit{Funding}: Advocate for funding mechanisms that explicitly support interdisciplinary HPSC efforts. Agencies should issue targeted calls that value not only scientific outcomes, but also software infrastructure, community building, and education. Projects at the intersection of statistical methodology and HPC infrastructure must be recognized as central to scientific innovation, not peripheral.
\end{enumerate}

\begin{figure*}
\centering
    \includegraphics[width=0.99\textwidth]{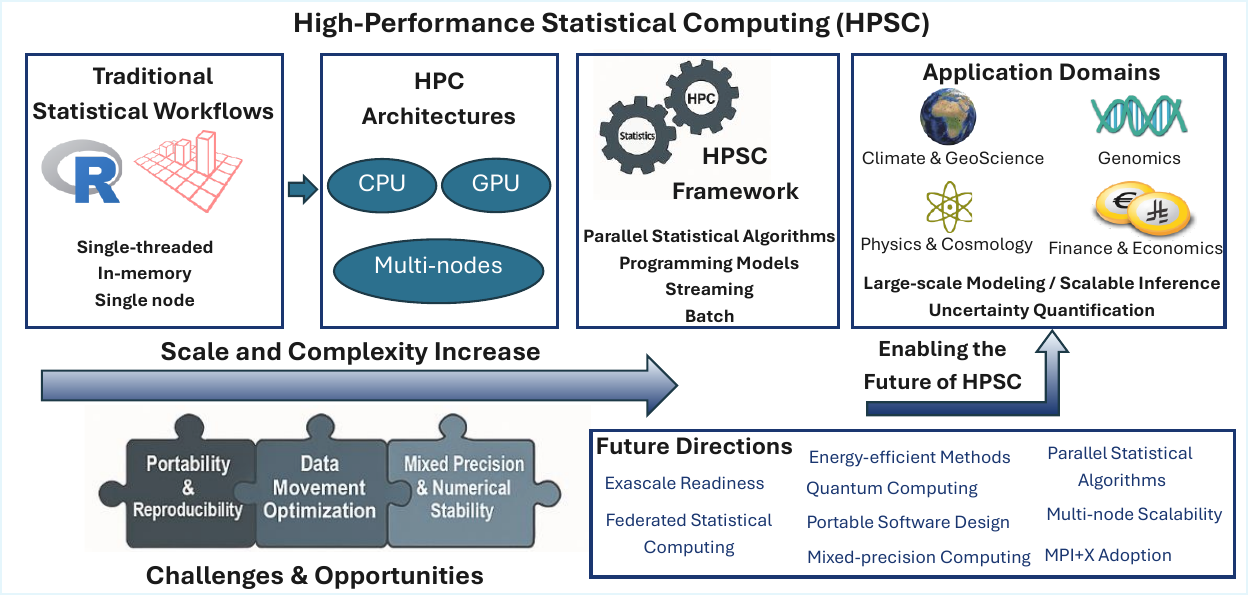} % Change filename
    \captionof{figure}{The evolution of HPSC: transitioning from traditional, single-threaded R workflows to scalable, multi-node architectures empowered by parallel statistical algorithms, HPC programming models, and emerging technologies such as federated computing, quantum acceleration, and mixed-precision methods.}
    \label{fig:hpsc}
\end{figure*}

A successful HPSC ecosystem requires shared responsibility. HPC scientists must focus on usability, abstraction, and system scalability, as well as designing software that lowers the barrier to entry while maximizing performance across various architectures. Statisticians should refine their existing and future methods to facilitate parallelism and distributed execution, adopt new paradigms, and promote reproducibility in HPC contexts. Both groups demonstrate a commitment to collaborating to promote diversity, equity, and inclusion in their training, outreach, and hiring practices.

The future of HPSC lies in this collective effort. As illustrated in Figure~\ref{fig:hpsc}, the HPSC landscape aims to change traditional single-threaded, in-memory workflows in existing statistical code powered by languages such as R with scalable, multi-node execution models powered by HPC architectures, parallel statistical algorithms, and emerging technologies such as federated computing, mixed-precision methods, and quantum acceleration. These advances support applications ranging from climate science to finance, driven by the growing scale of data and increasing computational complexity. The community can push the boundaries of what is statistically and computationally possible by combining the algorithmic depth of statistics with the computational power of HPC.
\newpage
\bibliographystyle{WileyNJD-APA}
\bibliography{sample}

\end{document}